\documentclass[sigconf,nonacm]{acmart}

\usepackage{booktabs}
\usepackage{array}
\usepackage{makecell}

\newcolumntype{L}[1]{>{\raggedright\arraybackslash}m{#1}}
\newcommand{\code}[1]{\mbox{\kern0.12em\texttt{#1}\kern0.12em}}

\begin{document}

\setcopyright{none}

\title{HINT: Toward an Executable Hardware-Intent Representation Layer for LLM-Driven RTL Generation}

\author{Tairan Cheng}
\affiliation{%
  \institution{The Chinese University of Hong Kong}
  \city{Hong Kong}
  \country{China}
}

\author{Yi Liu}
\affiliation{%
  \institution{The Chinese University of Hong Kong}
  \city{Hong Kong}
  \country{China}
}

\author{Dongsheng Zuo}
\affiliation{%
  \institution{The Chinese University of Hong Kong}
  \city{Hong Kong}
  \country{China}
}

\author{Zhengyuan Shi}
\affiliation{%
  \institution{The Chinese University of Hong Kong}
  \city{Hong Kong}
  \country{China}
}

\author{Hongji Zhang}
\affiliation{%
  \institution{The Chinese University of Hong Kong}
  \city{Hong Kong}
  \country{China}
}

\author{Xiangfei Hu}
\affiliation{%
  \institution{Southeast University}
  \city{Nanjing}
  \country{China}
}

\author{Maoshuo He}
\affiliation{%
  \institution{Southeast University}
  \city{Nanjing}
  \country{China}
}

\author{Hao Yan}
\affiliation{%
  \institution{Southeast University}
  \city{Nanjing}
  \country{China}
}

\author{Qiang Xu}
\affiliation{%
  \institution{The Chinese University of Hong Kong}
  \city{Hong Kong}
  \country{China}
}

\renewcommand{\shortauthors}{Cheng et al.}

\begin{abstract}
Generating implementation-quality RTL with large language models (LLMs) remains difficult because direct generation must resolve microarchitecture while simultaneously producing and debugging low-level code. We present HINT, an executable hardware-intent intermediate representation layer between behavioral specifications or executable oracles and RTL. HINT makes RTL-relevant microarchitecture explicit, supports pre-RTL checking, and supplies explicit RTL-lowering obligations. We evaluate HINT using both a minimal single-agent flow and a full staged workflow. Across seven operator cases, the HINT-mediated route, with no post-synthesis QoR refinement, produces contract-compliant synthesizable RTL on 7/7 cases; Direct C2RTL and C2HLSC apply to five cases and succeed on 5/5 and 1/5, respectively. Under matched Design Compiler synthesis, HINT reduces area by 5.0\%--26.2\% relative to five manual RTL implementations and by 8.9\%--86.1\% relative to five accepted Direct C2RTL results. RealBench AES and SDC, together with a Vortex VPU synthesizing to 561.67k~$\mu\mathrm{m}^2$, further demonstrate specification-driven, protocol-rich, memory-rich, and hierarchical designs. In the controlled operator study, the HINT-mediated route shows better observed convergence and avoids the severe implementation-quality degradation seen in several direct-generation results.
\end{abstract}


\maketitle

\section{Introduction}
\label{sec:introduction}

Large language models (LLMs) are rapidly improving in code generation, long-horizon reasoning, and tool-assisted workflows, making their use in front-end hardware design increasingly plausible~\cite{LLM4EDA,pan2025llm_eda_survey,abdollahi2025hdv_llm_review}. For example, HORIZON reports 100\% completion on the compact VerilogEval and RTLLM benchmarks~\cite{verilogeval,rtllm,yu2026horizon}, showing that current workflows handle compact RTL tasks well. The harder question is whether LLMs can generate correct, synthesizable RTL for designs with long specifications, interconnected state, and implementation-quality targets.

Hardware generation is not simply software generation in Verilog syntax. RTL remains far less represented in public corpora: in The Stack v2~\cite{lozhkov2024starcoder2stackv2}, CodeV reports 80.6M Python data points but only 1.91M Verilog data points~\cite{codev}. More fundamentally, behavioral intent rarely fixes the microarchitecture that determines correctness and quality of results (QoR). Direct RTL generation must infer it while simultaneously realizing and debugging ports, resets, clocks, and signal-level behavior. As designs grow, verification feedback therefore mixes architecture-level mistakes with RTL coding errors, making failures harder to localize and implementation QoR less stable.

This coupling exposes a representation gap. A useful intermediate representation layer should make RTL-relevant architectural decisions explicit and checkable, yet remain above cycle- and signal-level realization. We therefore introduce \emph{Hardware INTent representation layer (HINT)}, a specification-backed, executable representation between behavioral design intent and RTL. HINT makes typed transaction boundaries, owned state, bounded progress, control/datapath organization, resource commitments, and RTL-lowering obligations explicit, turning RTL generation from open-ended structure inference into constrained realization. Our evaluated instantiation expresses these hardware commitments in a restricted C-like executable form. HINT may be constructed from a textual specification, an executable C/C++ model, or both, so executable models strengthen verification without becoming a required input.

The separation is particularly useful for complex designs. HINT is executed and repaired at transaction level before RTL exists, allowing functional and state-transition problems to be debugged with conventional compiler diagnostics, software debugging tools, and oracle comparison instead of immediately entering signal-level simulation. Once accepted, its explicit microarchitecture commitments become RTL-lowering obligations. By keeping these high-impact decisions visible before low-level correctness repair dominates generation, HINT directs model effort toward the part of RTL coding that most strongly determines implementation QoR, without requiring an external QoR predictor or repeated post-synthesis search.

We evaluate HINT at two scales. In a controlled study of seven architecture-bearing operators, HINT covers and succeeds on 7/7 cases; Direct C2RTL and C2HLSC apply to five cases and succeed on 5/5 and 1/5, respectively. Under matched synthesis, it uses 5.0\%--26.2\% less area than five manual implementations and 8.9\%--86.1\% less than five accepted Direct C2RTL results.

We then apply the full staged workflow to RealBench AES and SDC~\cite{realbench} and a connected Vortex-derived non-floating-point VPU~\cite{tine2021vortex}, spanning specification-driven IP, protocols, memory, and hierarchy. All pass their declared checks, and the VPU synthesizes to 561.67k~$\mu\mathrm{m}^2$.

\noindent\textbf{Contributions.}
This work contributes (1) a semantically defined, executable hardware-intent intermediate representation layer that exposes high-impact structural decisions in RTL generation; (2) a HINT-centered methodology that separates transaction-level construction and debugging from RTL realization governed by explicit lowering obligations; and (3) controlled and cross-scale evaluations spanning operators, RealBench IP, and a connected repository-derived VPU subsystem.

\raggedbottom
\section{Related Work}
\label{sec:related_work}

\subsection{LLM-Based RTL Generation and Agentic Flows}

Recent work has improved direct RTL generation through model specialization, discriminative guidance, automated feedback, and candidate search. BetterV introduces controlled generation with discriminative guidance, RTLCoder develops an RTL-specialized open model, AutoVCoder organizes automated generation and repair, and MCTS-RTL searches over candidate implementations~\cite{pei2024betterv,fang2025rtlcoder,autovcoder,mcts_rtl}. AutoVeriFix uses Python reference model and coverage-refined tests to guide iterative Verilog correction through simulation mismatches~\cite{tan2026autoverifix}. Its Python model serves as debugging oracle rather than an design commitment. Agentic workflows further improve planning, iteration, and tool use. These methods strengthen the model or its coding process, but their target artifact remains RTL, so architectural commitment and low-level realization are still performed within the same generation stage. CktEvo instead benchmarks repository-level evolution of existing RTL artifacts~\cite{shi2026cktevo}; it addresses a complementary downstream setting rather than the pre-RTL representation studied here.

Benchmark development has made this progress measurable. VerilogEval contains 156 short HDLBits exercises whose human- and machine-authored references average only \textbf{15.8} and \textbf{13.9} RTL lines, respectively, with no submodule hierarchy; RTLLM contains 30 arithmetic and logic designs with median/mean/maximum sizes of 52/86/518 lines~\cite{verilogeval,rtllm,realbench}. HORIZON reports 100\% completion of both through a hands-free iterative tool-feedback loop~\cite{yu2026horizon}, showing that these compact suites are becoming less discriminative for strong workflows. More realistic suites retain a substantial gap: RealBench reports 13.3\% formal@1 at module level and 0\% at system level for o1-preview~\cite{realbench}, while ChipBench reports 30.74\% for its strongest evaluated model~\cite{yu2026chipbenchnextstepbenchmarkevaluating}.

\noindent\textbf{Benchmark choice.}
We therefore do not repeat VerilogEval or RTLLM. Their compact, mostly flat tasks remain useful tests of basic RTL coding. We instead focus on cases that expose more of the hardware-organization choices represented by HINT. Our operator cases add architecture-bearing datapath and control organization; RealBench adds specification- and protocol-rich IP; and the connected Vortex VPU adds repository context and hierarchical integration. These cases test whether an intermediate representation layer remains useful beyond short isolated modules.

\subsection{High-Level and Intermediate Hardware Representations}

C-based high-level synthesis (HLS) compiles behavioral C/C++ into RTL through scheduling, allocation, binding, and related optimizations~\cite{cong2011hls_fpga,nane2016hls_survey}. More recently, C2HLSC uses an LLM to rewrite reference C into Catapult-compatible HLS C, demonstrating that generation at a higher abstraction level can substantially improve feasibility~\cite{c2hlsc}. Its output, however, remains behavioral HLS C: scheduling, binding, control generation, and much of the resulting hardware organization are selected by the HLS tool. HINT instead requires the major architecture commitments to be explicit before RTL lowering.

SystemC spans a different abstraction range. Transaction-level modeling supports system modeling and virtual prototyping without committing to detailed implementation timing~\cite{cai2003tlm_overview,donlin2004tlm_flows}. Synthesizable or cycle-accurate SystemC can encode detailed hardware behavior, but does so through clocked processes, sensitivity, explicit waits, and cycle-level concurrency~\cite{accellera_systemc_synth_2016}. It therefore approaches the realization burden that HINT intentionally defers.

Hardware compiler intermediate representations address different stages of the flow. Calyx represents a supplied accelerator architecture through structural components and an explicit control schedule, while FIRRTL supports circuit transformation and lowering after hardware-generator elaboration~\cite{nigam2021calyx,li2016firrtl}. MLIR and CIRCT provide broader reusable infrastructure for multi-level dialects and for transforming and lowering hardware representations~\cite{lattner2021mlir,eldridge2021mlir_hardware}. CPPL uses a Python prompt-programming frontend to predeclare module interfaces, hierarchy, and instance connections, and asks the LLM to complete a statically checked JSON circuit body for CIRCT lowering~\cite{yin2026cppl}. Its checks establish circuit legality, including widths and structural bindings, while behavioral intent remains in natural-language descriptions and functional correctness is evaluated by external tests.

These systems provide compiler machinery after a frontend supplies hardware or circuit structure. HINT targets the preceding LLM-facing architecture-authoring stage, where architecture commitments are made explicit and executable before RTL realization.
\section{HINT Representation}
\label{sec:definition}

\noindent\textbf{Reference profile.}
HINT can support different executable forms for expressing the hardware commitments summarized in Table~\ref{tab:HINT_norm}. Our experiments instantiate HINT with an engineer-defined, case-independent restricted C-like reference profile. The checker-enabled full workflow enforces the reference-profile specification.

HINT is IR-like in role: it is an LLM-authorable, executable, and tool-checkable intermediate artifact whose accepted semantics constrain downstream realization. Unlike a conventional compiler IR built around deterministic passes, HINT targets LLM-friendly architecture authoring and leverages the flexibility of agentic construction and lowering; its accepted architectural commitments bound the resulting implementation variation.

\begin{table}[t]
  \centering
  \caption{Required HINT-layer obligations.}
  \label{tab:HINT_norm}
  \footnotesize
  \renewcommand{\arraystretch}{1.05}
  \setlength{\tabcolsep}{3pt}
  \begin{tabular}{L{0.28\columnwidth} L{0.65\columnwidth}}
    \toprule
    \textbf{Layer obligation} & \textbf{Defined hardware meaning} \\
    \midrule
    Program and object model & Units, persistent child/helper instances, hierarchy, and explicit ownership. \\
    Types, interfaces, behavior & Fixed-width objects, typed transaction boundaries, operators, and memory/buffer semantics. \\
    State, control, execution & Current/next state, reset/commit, bounded progress, and combinational/FSM/phase/counter/pipeline models. \\
    Resource and QoR model & Sharing, replication, reuse scope, staging, storage, and QoR-relevant annotations. \\
    RTL-lowering obligation & Interface, state, datapath, schedule, and resource commitments carried into RTL. \\
    \bottomrule
  \end{tabular}
\end{table}

\noindent\textbf{Executable hardware semantics.}
A HINT design is a hierarchy of persistent \emph{units} and instances, not a collection of software subroutine activations. Each unit has typed boundary objects, owns its carried state, and exposes exactly one step entry point under its execution model:
\begin{equation}
  (o_t,s_{t+1})=\mathrm{step}_{\mathcal H}(i_t,s_t),
  \label{eq:hint_step}
\end{equation}
where $i_t$, $o_t$, and $s_t$ denote input, output, and unit-owned architectural state. Here $t$ indexes an abstract advancement under the selected execution model, not necessarily an RTL clock cycle. A step observes current state, produces outputs and next state, and commits updates at its boundary. HINT can therefore be executed against a behavioral or transaction-level oracle before RTL exists.

\noindent\textbf{Module/IP-level granularity.}
HINT makes canonical microarchitecture commitments explicit without becoming cycle-accurate or signal-level code. It may commit to latency, staging, or resource schedules where required, but need not encode per-cycle signal assignments or RTL process structure. A HINT unit denotes a module/IP-level hardware block, and an instance denotes a persistent child block rather than a function call. Hierarchical designs are expressed by composing such units, without making HINT another RTL syntax.

\noindent\textbf{Canonical hardware form.}
In our reference profile, a conforming artifact is statically bounded, excludes dynamic allocation, recursion, unbounded behavior, and hidden side effects, and carries only architecture-relevant state. ``Canonical'' requires the selected architecture to remain visible through semantic objects; it does not require unique identifiers or a unique textual rendering. In the checker-enabled cross-scale workflow, the case-independent HINT Lint Checker validates permitted forms and tags, typed boundaries and references, state ownership and commit discipline, bounded progress, and instance/resource cardinality.

\begin{figure}[t]
  \centering
  \includegraphics[width=\columnwidth]{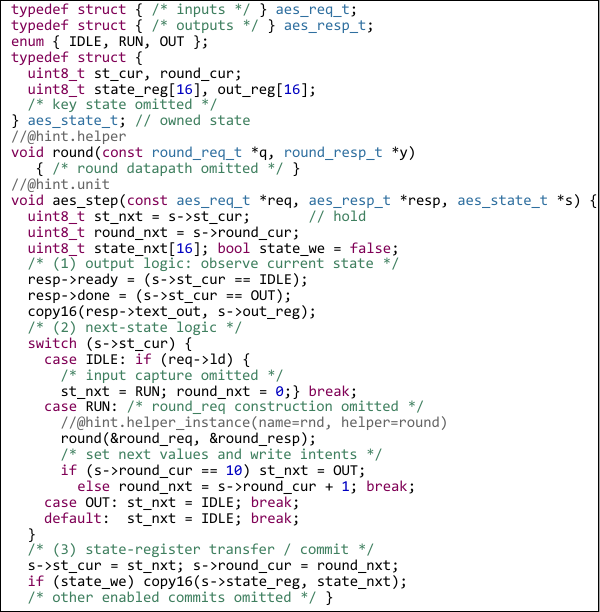}
  \caption{Canonical reference-profile excerpt for a shared-round AES unit.}
  \label{fig:hint_aes_excerpt}
\end{figure}

Figure~\ref{fig:hint_aes_excerpt} shows these properties for a shared-round AES unit. For space, the figure omits the byte-level AES implementation logic; the complete HINT artifact includes this logic, and executable acceptance checks its correctness rather than treating it as an uninterpreted operation.

The request, response, and state objects separate typed boundaries from owned state. Within the state object, the control mode and round index encode bounded progress; outputs observe current state, next-state logic defaults to hold, and owned updates occur only at commit. The named round instance declares one datapath reused across all rounds. Thus, the skeleton fixes control, progress, and resource organization without specifying clocks, signals, or RTL process style.

\noindent\textbf{RTL-lowering obligations.}
HINT is an executable architecture contract, not a prompt-level plan or an RTL template. Its accepted interface, state, schedule, and resource commitments become explicit obligations for RTL lowering, while local coding structure may vary. HINT--RTL comparison checks observable behavior and transaction timing rather than claiming a formal proof of internal structural equivalence. Architectural changes require renewed HINT acceptance.
\section{Methodology}
\label{sec:methodology}

\subsection{Staged Architecture Construction and RTL Lowering}

\begin{figure*}[t]
  \centering
  \includegraphics[width=\textwidth]{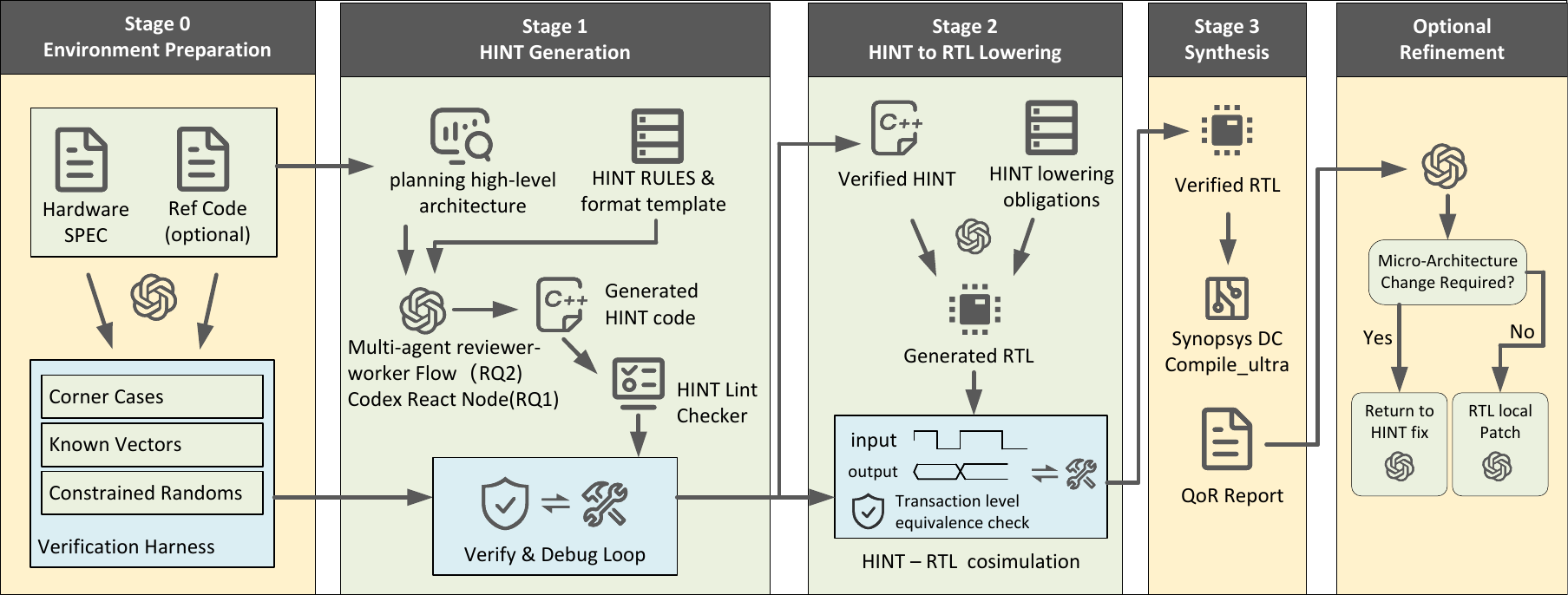}
  \caption{Full HINT-centered flow used for cross-scale generation. The controlled operator study uses the reduced sequential workflow described in Sec.~\ref{subsec:setup}.}
  \label{fig:workflow}
\end{figure*}

Figure~\ref{fig:workflow} shows the execution order. The HINT layer can be independent of any particular agent framework. In our experiments, the operator study uses one sequential Codex ReAct node with fixed feedback, whereas RealBench and Vortex add canonical-form checking, environment construction, architecture review, and, when needed, ownership partitioning and integration. Both workflow configurations accept HINT through transaction-level checking before RTL lowering.

\noindent\textbf{Stage 0---Task and verification preparation.}
Inputs are a design specification and, when available, an executable reference model or native tests; C/C++ serves as an oracle rather than a required HINT input. For raw benchmarks or repositories, the environment agent resolves scope and dependencies, extracts interface/transaction/timing contracts, partitions by hardware ownership when needed, and builds the verification environment. Reference-model cases combine known vectors, directed corners, and constrained-random transactions; RealBench reuses native tests or locked test collateral derived from them. These assets and acceptance rules are frozen before generation and cannot be modified by design agents.

\noindent\textbf{Formal-checking scope.}
For the 8-bit \code{Hif8\_mul} case, we additionally use Synopsys Formality~\cite{synopsys_formality} to check netlist-level equivalence between the HINT-lowered and engineer-written implementations. The remaining designs are outside the scope of our current formal tools and are accepted through transaction-level regression/co-simulation.

\noindent\textbf{Stage 1---HINT generation and acceptance.}
The agent first determines the microarchitecture and encodes it as a HINT artifact using the reference profile in Sec.~\ref{sec:definition}. In the checker-enabled full workflow, the case-independent HINT Lint Checker validates the canonical-form obligations described above. It reports structural violations but neither proposes a microarchitecture nor estimates QoR. In both workflow configurations, the executable HINT artifact is evaluated using the Stage-0 reference or native checking environment with the same frozen tests or transactions, comparing outputs and boundary events at transaction level. Once the applicable checks pass, the accepted HINT artifact fixes the microarchitecture and governs RTL lowering.

\noindent\textbf{Stage 2---RTL lowering and acceptance.}
The lowering obligations guide realization of the accepted HINT artifact as synthesizable RTL without re-inferring the organization from behavioral code. Using the same frozen tests or transactions, the checking environment compares HINT and RTL for observable data and timing and, where available, also compares them with the reference model or original RTL; otherwise, the predeclared design specification supplies target latency. Realization errors may be repaired locally, but architectural changes return to Stage~1. Only RTL passing all applicable checks proceeds.

\noindent\textbf{Stage 3---Synthesis and QoR reporting.}
Accepted RTL is synthesized to report area, critical-path timing, and area--delay product (ADP). Optional synthesis feedback may request a local RTL repair or, if the architecture must change, a return to Stage~1; either path must repeat the applicable conformance and functional gates. For RQ1, the controlled HINT route stops at the first successful synthesis after correctness acceptance and does not use this optional refinement.

\noindent\textbf{Failure localization.}
The gates localize failures: where enabled, HINT Lint flags conformance defects; HINT--reference/test mismatches expose transaction/state defects; HINT--RTL transaction mismatches expose behavioral or timing errors introduced during lowering; and synthesis reveals QoR pressure after correctness. Larger designs can thus be debugged with transaction traces before analysis moves to signal-level RTL.

\begin{table*}[!t]
  \centering
  \caption{Controlled operator results.}
  \label{tab:operator_results}
  \small
  \renewcommand{\arraystretch}{1.05}
  \setlength{\tabcolsep}{1.5pt}
  \begin{tabular*}{\textwidth}{@{\extracolsep{\fill}}l c c c c c@{}}
    \toprule
    \textbf{Case}
      & \makecell{\textbf{Target}\\\textbf{Latency}}
      & \makecell{\textbf{HINT route}\\Area/Timing/ADP/Debug Iter.}
      & \makecell{\textbf{Manual RTL}\\Area/Timing/ADP}
      & \makecell{\textbf{Direct C2RTL}\\Area/Timing/ADP/Debug Iter.}
      & \makecell{\textbf{C2HLSC flow}\\Area/Timing/ADP/Debug Iter.} \\
    \midrule
    Hif8\_mul (pilot) & 0 & \textbf{1.58k}/7.82/\textbf{12.3}/1
      & 2.14k/8.67/18.5 & NA & NA \\
    Karatsuba & 0 & \textbf{8.90k}/15.48/\textbf{137.8}/0
      & 9.38k/15.08/141.3 & NA & NA \\
    CORDIC & 3 & \textbf{11.10k}/15.89/\textbf{176.0}/1
      & 12.28k/22.24/273.2 & 54.41k/59.85/3.26k/3
      & 14.22k/16.48/234.4/11 \\
    SHA256 & 66 & \textbf{34.18k}/17.15/\textbf{586.1}/5
      & 42.75k/16.84/720.0 & 60.33k/19.59/1.18k/8
      & 134.19k/19.37/2.60k/17$^{\star}$ \\
    MD5 & 68 & \textbf{37.25k}/32.44/\textbf{1.21k}/3
      & 39.66k/32.05/1.27k & 208.48k/614.24/128.06k/5
      & 52.64k/30.60/1.61k/9$^{\star}$ \\
    AES$_{\mathrm{C2HLSC}}$ & 14 & \textbf{68.14k}/6.18/\textbf{421.1}/0
      & NA & 489.85k/39.85/19.52k/0 & NC \\
    DES$_{\mathrm{C2HLSC}}$ & 18 & \textbf{10.58k}/6.98/\textbf{73.9}/1
      & NA & 11.61k/6.71/77.9/1 & NC \\
    \midrule
    \textbf{Accepted/applicable} & -- & 7/7 & 5/5 & 5/5 & 1/5 \\
    \bottomrule
  \end{tabular*}
  \vspace{1pt}

  \begin{minipage}{\textwidth}
    \footnotesize\raggedright
    \textit{Status:} NA denotes an inapplicable route. NC denotes failure within budget to obtain a contract-compliant result that can be mapped by Design Compiler. Debug Iter. counts correctness-repair revisions. $^{\star}$ denotes a C2HLSC output that misses the specified latency; its comparable-scale QoR is shown for context and excluded from strict QoR claims.
  \end{minipage}
\end{table*}

\subsection{QoR Comparison Protocol}
\label{subsec:ppa}

For each case, all accepted generated RTL and all available original RTL are synthesized using the same technology library, Synopsys Design Compiler \code{compile\_ultra} flow~\cite{synopsys_design_compiler}, scripts, optimization settings, timing constraints, and reporting conventions. Comparisons are therefore made within a case rather than across unrelated timing regimes. Clocked operator cases use a fixed 1000-ns synthesis constraint, whereas the RealBench and VPU cases use predeclared design-specific constraints; all compared implementations within a case share the same frozen constraint.

Total cell area is the primary metric for comparing mapped logic complexity under that fixed regime, preventing one route from obtaining a smaller design merely through a looser clock target. We nevertheless report critical-path delay separately to expose timing degradation, and report area--delay product, $\mathrm{ADP}=A_{\mathrm{total}}\!\times\!T_{\mathrm{cp}}$, as a supplementary combined metric following prior LLM-based hardware evaluation~\cite{shi2026cktevo}. ADP does not replace the separate area and timing results. Available original RTL is treated as a same-flow implementation reference, not as a globally optimized human upper bound. Generated routes must satisfy their frozen functional, latency, and initiation-interval contracts where applicable.

\noindent\textbf{Artifact release.}
Experimental artifacts are available at \url{https://anonymous.4open.science/r/anonymous-project-612D/}. The repository is for the review process, and the complete release with HINT SPEC will be made upon publication.
\section{Experiments}
\label{sec:evaluation}

\subsection{Experimental Design and Setup}
\label{subsec:setup}

We answer two questions. \emph{RQ1} asks whether replacing direct RTL emission with the HINT-mediated construction, acceptance, and lowering route improves correctness convergence and implementation QoR when orchestration is intentionally minimized. \emph{RQ2} asks whether the HINT approach remains usable for specification-driven IP, protocol-rich control, and a repository-derived hierarchical subsystem. The operator study provides controlled route-level evidence; RealBench and Vortex provide feasibility and scale evidence and are not pooled into the RQ1 comparison.

\noindent\textbf{Flow assignment.}
All runs use OpenAI's GPT-5.3-Codex~\cite{openai_gpt53codex_2026}. RQ1 uses one sequential Codex CLI ReAct run per route with a case-independent trigger and no case-specific architecture or repair guidance. The HINT route inserts construction of a pre-RTL artifact, while Direct C2RTL uses the same node to emit and revise RTL directly. The model, the frozen specification and contracts, any available oracle, the locked verifier, the correctness budget, and the orchestration remain fixed; RQ1 excludes the checker and multi-agent system used only in the RQ2 full workflow. HINT receives the reference-profile specification, while C2HLSC follows its reproduced Catapult protocol~\cite{c2hlsc}.

\noindent\textbf{Route-level ablation and convergence scope.}
The HINT-mediated route is an integrated representation workflow requiring construction, executable acceptance, and RTL lowering across multiple agentic stages; its appropriate ablation unit is therefore the route rather than individual components. RQ1 treats this route as a pluggable ablation unit: inserting it gives the HINT route, while removing it yields the matched Direct C2RTL route. To prevent the additional HINT stages from conferring a search advantage, they share the route-level correctness budget without a stage reset, and HINT is frozen at its first successful synthesis, whereas Direct C2RTL may continue post-synthesis QoR refinement. The comparison is thus conservative for HINT. ReAct revisions are adaptive rather than independent \emph{pass@k} samples; counting them as such would mischaracterize dependent repairs as fresh generations. In this agent setting with a strong base model and stateful tool feedback, we therefore characterize observed convergence by within-budget acceptance and correctness-repair revisions (Debug Iter.).

A non-corrective post-freeze source audit found that each lowered RTL retained the control, schedule, and resource organization declared in HINT. RQ2 uses the checker-enabled full workflow.

\noindent\textbf{RQ1 Guidance and iteration fairness.}
All routes share one 20-revision correctness budget. The controller returns only raw compiler, testbench, or synthesis diagnostics, and the Codex ReAct node chooses every diagnosis and edit without human RTL/HINT edits or candidate selection. Post-synthesis QoR refinement steps for Direct C2RTL and C2HLSC are not included in the reported correctness-debug count.

\noindent\textbf{Targets and synthesis.}
Target latency and initiation interval (II) are extracted from reference RTL when available and otherwise fixed manually before generation. For each case, the design specification and targets are then frozen and shared unchanged by all applicable routes. Accepted RTL is synthesized with SKY130~\cite{skywater_pdk_github} under Sec.~\ref{subsec:ppa}. Each comparison shares the same library, scripts, constraints, and reporting convention, and only contract-satisfying outputs enter strict QoR comparisons.

\subsection{Controlled Operator Study (RQ1)}

\noindent\textbf{Cases and comparison coverage.}
RQ1 covers \code{Hif8\_mul}~\cite{luo2024hifloat8}, Karatsuba, CORDIC~\cite{cordic}, SHA256, MD5, and the C2HLSC AES/DES tasks~\cite{c2hlsc}. HINT covers all seven; manual RTL covers five, and Direct C2RTL/C2HLSC apply to five. \code{Hif8\_mul} lacks reference C, Karatsuba's C model is only a multiplication expression and omits the intended recursive architecture, and AES/DES lack manual RTL; Table~\ref{tab:operator_results} reports all applicable routes without counting NA as failure. We apply one width-only lint correction to Karatsuba manual RTL. \code{Hif8\_mul} is a pilot in which an engineer authored the HINT artifact and Codex mechanically lowered it to RTL.

Target latency is reported in cycles, area in $\mu\mathrm{m}^2$, critical-path timing in ns, and ADP in $10^3\,\mu\mathrm{m}^2\!\cdot\!\mathrm{ns}$; suffix ``k'' denotes $\times 10^3$, and latency 0 denotes a combinational contract.

\noindent\textbf{Correctness and convergence.}
HINT produces contract-compliant, synthesizable RTL on all \textbf{7/7} cases. On the five cases commonly applicable to all three generated routes, HINT and Direct C2RTL each succeed on \textbf{5/5}, compared with \textbf{1/5} for C2HLSC.

\noindent\textbf{Implementation QoR.}
Under matched synthesis constraints, HINT reduces area by \textbf{5.0\%--26.2\%} relative to the five manual implementations and improves ADP in all five comparisons. Relative to the five accepted Direct C2RTL results, HINT reduces area by \textbf{8.9\%--86.1\%} and improves ADP in every comparison. CORDIC is the only strictly accepted C2HLSC comparison, where HINT is 22\% smaller and slightly faster. Each reported HINT result corresponds to the first successful synthesis after correctness acceptance, without synthesis-driven candidate selection or QoR refinement.

\noindent\textbf{Why HINT may benefit QoR.}
HINT separates high-impact microarchitectural decisions from low-level generation details and carries them into RTL through its lowering obligations, which can improve cross-case QoR consistency. The resulting organization can differ substantially from manual RTL but does not guarantee an optimal implementation.

\begin{figure}[t]
  \centering
  \includegraphics[width=\columnwidth]{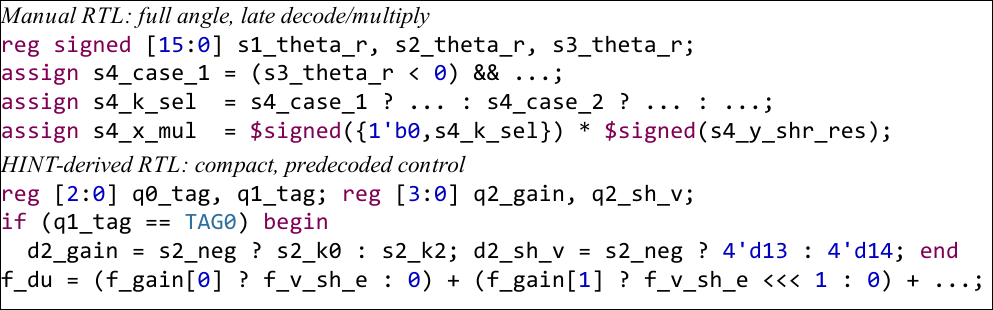}
  \caption{CORDIC example comparing two functionally matched RTL organizations.}
  \label{fig:cordic_qor_excerpt}
\end{figure}

Figure~\ref{fig:cordic_qor_excerpt} gives a concrete example of this structural difference: manual RTL carries full angles into a late decode and uses variable small-constant multiplication, whereas HINT-derived RTL carries compact tags, predecodes control, and exposes compensation as shift--add logic. Under the same synthesis flow, combinational area is nearly identical, while the HINT-derived organization reduces sequential cells from 164 to 117 and noncombinational area by 28.4\%; total area falls from 12.28k to 11.10k~$\mu\mathrm{m}^2$ and the critical path from 22.24 to 15.89~ns.

\begin{table}[!t]
  \centering
  \caption{Cross-scale implementation results.}
  \label{tab:cross_scale}
  \small
  \renewcommand{\arraystretch}{1.03}
  \setlength{\tabcolsep}{1.5pt}
  \begin{tabular*}{\columnwidth}{@{\extracolsep{\fill}}l r r r r@{}}
    \toprule
    \textbf{Design/block}
      & \makecell{\textbf{HINT}\\\textbf{LOC}}
      & \makecell{\textbf{RTL}\\\textbf{LOC}}
      & \makecell{\textbf{Area}\\($\mu\mathrm{m}^2$)}
      & \makecell{\textbf{Timing}\\(ns)} \\
    \midrule
    RealBench AES & 407 & 550 & 59,501 & 5.77 \\
    RealBench SDC & 1,579 & 1,732 & 96,426 & 7.18 \\
    \midrule
    \textbf{Vortex VPU top} & \textbf{2,744} & \textbf{3,541}
      & \textbf{561,667} & \textbf{14.69} \\
    \hspace{0.6em}\code{vcfg} & 461 & 654 & 14,647 & 4.64 \\
    \hspace{0.6em}\code{vrf} & 358 & 248 & 177,245 & 11.49 \\
    \hspace{0.6em}\code{int\_exec} & 809 & 707 & 94,411 & 14.54 \\
    \hspace{0.6em}\code{mv\_mask} & 465 & 380 & 39,590 & 5.62 \\
    \hspace{0.6em}\code{vlsu} & 651 & 871 & 160,720 & 13.35 \\
    \bottomrule
  \end{tabular*}
  \vspace{1pt}
\end{table}

\begin{table}[!t]
  \centering
  \caption{Protocol- and repository-scale feature coverage.}
  \label{tab:cross_scale_features}
  \small
  \renewcommand{\arraystretch}{1.01}
  \setlength{\tabcolsep}{1.2pt}
  \begin{tabular}{@{}L{0.23\columnwidth} L{0.73\columnwidth}@{}}
    \toprule
    \textbf{Design/block} & \textbf{Module features} \\
    \midrule
    RealBench SDC & Communication- and control-intensive SD-card controller spanning Wishbone slave/master traffic, SD command/data protocols, DMA, descriptor/FIFO progress, CRC/status/interrupt state, dynamic clocking, and dual-edge-visible timing. \\
    \cmidrule(l){2-2}
    Vortex-derived VPU & Major protocol- and memory-rich vector component of the RISC-V-based Vortex GPU: a connected 128-bit non-floating-point subsystem with RVV-style configuration/CSR state, 32 architectural vector registers, SEW 8/16/32 integer/multiply/mask/move execution, masked commit, and unit-stride memory request/response handling. \\
    \bottomrule
  \end{tabular}
\end{table}

\subsection{Cross-Scale Studies (RQ2)}

\noindent\textbf{Cases and verification.}
We evaluate RealBench AES and SDC with their native environments~\cite{realbench}; E203 is omitted because its released CPU-top harness provides only short pin-level stimuli without processor workloads. For Vortex commit \code{fe8168b}~\cite{tine2021vortex}, a C++ extract frozen before generation produces 48 directed and 500 seeded-random root transactions covering configuration/CSR behavior, OPIVV/OPIVI/OPIVX/OPMVV/OPMVX families, zero and short lengths, and unit-stride memory widths 1/2/4. Tables~\ref{tab:cross_scale} and~\ref{tab:cross_scale_features} report implementation scale and feature coverage.

\noindent\textbf{RealBench results.}
Neither RealBench case provides reference C. Generated AES passes the native functional, reset, and exact-latency checks, while generated SDC passes its native protocol tests. Under the same synthesis flow, generated SDC uses 96.43k versus 99.62k~$\mu\mathrm{m}^2$ for the original RTL, while its critical path is 7.18 versus 5.11~ns. The two implementations thus show an area--timing trade-off.

\noindent\textbf{Vortex-derived non-floating-point VPU.}
During Stage~0, the VPU's non-floating-point scope is partitioned by hardware ownership into the five blocks in Table~\ref{tab:cross_scale}. For each block, both HINT and RTL pass co-simulation against the same pinned C++ oracle before recomposition; the physically connected design is then checked again at the top transaction boundary, so the evaluation does not reduce to five isolated module tests.

The connected VPU synthesizes to \textbf{561.67k~$\mu\mathrm{m}^2$}. Table~\ref{tab:cross_scale} shows the block-level breakdown; the relatively large VRF and VLSU areas reflect the absence of memory macros in our library, so their storage and buffering must be implemented with standard-cell registers.

\section{Conclusion}

This work presents HINT, an executable and checkable hardware-intent intermediate representation layer that makes high-impact microarchitectural decisions explicit before RTL realization and carries them as lowering obligations. In the controlled operator study, the HINT-mediated route shows better observed convergence than the generated baselines and produces consistently competitive area and timing QoR without post-synthesis QoR refinement; RealBench and the Vortex-derived VPU further demonstrate applicability to protocol-rich IP and connected hierarchy. By separating architecture construction from signal-level repair, HINT exposes high-impact implementation choices before RTL coding begins. Its layer obligations and reference profile provide a foundation for reusable intent artifacts and future checking, exploration, and lowering tools.
\bibliographystyle{ACM-Reference-Format}
\bibliography{sample-base}

@inproceedings{pei2024betterv,
  title={BetterV: Controlled Verilog Generation with Discriminative Guidance},
  author={Pei, Zehua and Zhen, Huiling and Yuan, Mingxuan and Huang, Yu and Yu, Bei},
  booktitle={International Conference on Machine Learning},
  pages={40145--40153},
  year={2024},
  organization={PMLR}
}

@article{LLM4EDA,
author = {He, Zhuolun and Pu, Yuan and Wu, Haoyuan and Qiu, Tairu and Yu, Bei},
title = {Large Language Models for EDA: Future or Mirage?},
year = {2025},
issue_date = {November 2025},
publisher = {Association for Computing Machinery},
address = {New York, NY, USA},
volume = {30},
number = {6},
issn = {1084-4309},
url = {https://doi.org/10.1145/3736167},
doi = {10.1145/3736167},
journal = {ACM Trans. Des. Autom. Electron. Syst.},
month = oct,
articleno = {90},
numpages = {53}
}

@ARTICLE{codev,
  author={Zhao, Yang and Huang, Di and Li, Chongxiao and Jin, Pengwei and Song, Muxin and Xu, Yinan and Nan, Ziyuan and Gao, Mingju and Ma, Tianyun and Qi, Lei and Pan, Yansong and Zhang, Zhenxing and Zhang, Rui and Zhang, Xishan and Du, Zidong and Guo, Qi and Hu, Xing},
  journal={IEEE Transactions on Computer-Aided Design of Integrated Circuits and Systems}, 
  title={CodeV: Empowering LLMs with HDL Generation through Multi-Level Summarization}, 
  year={2025},
  volume={},
  number={},
  pages={1-1},
  doi={10.1109/TCAD.2025.3604320}}

@INPROCEEDINGS{verilogeval,
  author={Liu, Mingjie and Pinckney, Nathaniel and Khailany, Brucek and Ren, Haoxing},
  booktitle={2023 IEEE/ACM International Conference on Computer Aided Design (ICCAD)}, 
  title={Invited Paper: VerilogEval: Evaluating Large Language Models for Verilog Code Generation}, 
  year={2023},
  volume={},
  number={},
  pages={1-8},
  doi={10.1109/ICCAD57390.2023.10323812}}

@misc{realbench,
      title={RealBench: Benchmarking Verilog Generation Models with Real-World IP Designs}, 
      author={Pengwei Jin and Di Huang and Chongxiao Li and Shuyao Cheng and Yang Zhao and Xinyao Zheng and Jiaguo Zhu and Shuyi Xing and Bohan Dou and Rui Zhang and Zidong Du and Qi Guo and Xing Hu},
      year={2025},
      eprint={2507.16200},
      archivePrefix={arXiv},
      primaryClass={cs.LG},
      url={https://arxiv.org/abs/2507.16200}, 
}

@misc{autovcoder,
      title={AutoVCoder: A Systematic Framework for Automated Verilog Code Generation using LLMs}, 
      author={Mingzhe Gao and Jieru Zhao and Zhe Lin and Wenchao Ding and Xiaofeng Hou and Yu Feng and Chao Li and Minyi Guo},
      year={2024},
      eprint={2407.18333},
      archivePrefix={arXiv},
      primaryClass={cs.AR},
      url={https://arxiv.org/abs/2407.18333}, 
}

@article{c2hlsc,
author = {Collini, Luca and Garg, Siddharth and Karri, Ramesh},
title = {C2HLSC: Leveraging Large Language Models to Bridge the Software-to-Hardware Design Gap},
year = {2025},
issue_date = {November 2025},
publisher = {Association for Computing Machinery},
address = {New York, NY, USA},
volume = {30},
number = {6},
issn = {1084-4309},
url = {https://doi.org/10.1145/3734524},
doi = {10.1145/3734524},
journal = {ACM Trans. Des. Autom. Electron. Syst.},
month = oct,
articleno = {96},
numpages = {24}
}

@article{mcts_rtl,
  title        = {Make Every Move Count: {LLM}-based High-Quality RTL Code Generation Using {MCTS}},
  author       = {DeLorenzo, Matthew and Basak Chowdhury, Animesh and Gohil, Vasudev and Thakur, Shailja and Karri, Ramesh and Garg, Siddharth and Rajendran, Jeyavijayan},
  journal      = {arXiv preprint arXiv:2402.03289},
  year         = {2024},
  eprint       = {2402.03289},
  archivePrefix= {arXiv},
  primaryClass = {cs.LG}
}

@misc{lozhkov2024starcoder2stackv2,
      title={StarCoder 2 and The Stack v2: The Next Generation}, 
      author={Anton Lozhkov and Raymond Li and Loubna Ben Allal and Federico Cassano and Joel Lamy-Poirier and Nouamane Tazi and Ao Tang and Dmytro Pykhtar and Jiawei Liu and Yuxiang Wei and Tianyang Liu and Max Tian and Denis Kocetkov and Arthur Zucker and Younes Belkada and Zijian Wang and Qian Liu and Dmitry Abulkhanov and Indraneil Paul and Zhuang Li and Wen-Ding Li and Megan Risdal and Jia Li and Jian Zhu and Terry Yue Zhuo and Evgenii Zheltonozhskii and Nii Osae Osae Dade and Wenhao Yu and Lucas Krauß and Naman Jain and Yixuan Su and Xuanli He and Manan Dey and Edoardo Abati and Yekun Chai and Niklas Muennighoff and Xiangru Tang and Muhtasham Oblokulov and Christopher Akiki and Marc Marone and Chenghao Mou and Mayank Mishra and Alex Gu and Binyuan Hui and Tri Dao and Armel Zebaze and Olivier Dehaene and Nicolas Patry and Canwen Xu and Julian McAuley and Han Hu and Torsten Scholak and Sebastien Paquet and Jennifer Robinson and Carolyn Jane Anderson and Nicolas Chapados and Mostofa Patwary and Nima Tajbakhsh and Yacine Jernite and Carlos Muñoz Ferrandis and Lingming Zhang and Sean Hughes and Thomas Wolf and Arjun Guha and Leandro von Werra and Harm de Vries},
      year={2024},
      eprint={2402.19173},
      archivePrefix={arXiv},
      primaryClass={cs.SE},
      url={https://arxiv.org/abs/2402.19173}, 
}

@manual{accellera_systemc_synth_2016,
  title        = {SystemC Synthesizable Subset Language Reference Manual, Version 1.4.7},
  author       = {{Accellera Systems Initiative}},
  year         = {2016},
  organization = {Accellera Systems Initiative},
  url          = {https://www.accellera.org/images/downloads/standards/systemc/SystemC_Synthesis_Subset_1_4_7.pdf}
}

@inproceedings{nigam2021calyx,
  author    = {Rachit Nigam and Samuel Thomas and Zhijing Li and Adrian Sampson},
  title     = {A Compiler Infrastructure for Accelerator Generators},
  booktitle = {Proceedings of the 26th ACM International Conference on Architectural Support for Programming Languages and Operating Systems (ASPLOS)},
  year      = {2021},
  pages     = {804--817},
  doi       = {10.1145/3445814.3446712},
  url       = {https://www.cs.cornell.edu/~asampson/media/papers/calyx-asplos2021.pdf}
}

@techreport{li2016firrtl,
  author      = {Patrick S. Li and Adam Izraelevitz and Jonathan Bachrach},
  title       = {Specification for the FIRRTL Language},
  institution = {EECS Department, University of California, Berkeley},
  number      = {UCB/EECS-2016-9},
  year        = {2016},
  url         = {https://www2.eecs.berkeley.edu/Pubs/TechRpts/2016/EECS-2016-9.pdf}
}

@article{luo2024hifloat8,
  author  = {Yuanyong Luo and Zhongxing Zhang and Richard Wu and Hu Liu and Ying Jin and Kai Zheng and Minmin Wang and Zhanying He and Guipeng Hu and Luyao Chen and Tianchi Hu and Junsong Wang and Minqi Chen and Dmitry Mikhaylov and Vladimir Korviakov and Maxim Bobrin and Yuhao Hu and Guanfu Chen and Zeyi Huang},
  title   = {Ascend HiFloat8 Format for Deep Learning},
  journal = {arXiv preprint arXiv:2409.16626},
  year    = {2024},
  url     = {https://arxiv.org/abs/2409.16626}
}

@misc{skywater_pdk_github,
  author       = {{Google and SkyWater Technology}},
  title        = {Open Source Process Design Kit for the SkyWater SKY130 Process Node},
  year         = {2026},
  howpublished = {\url{https://github.com/google/skywater-pdk}},
  note         = {Accessed 2026-04-10}
}

@misc{openai_gpt53codex_2026,
  author       = {{OpenAI}},
  title        = {GPT-5.3-Codex Model},
  year         = {2026},
  howpublished = {\url{https://developers.openai.com/api/docs/models/gpt-5.3-codex}},
  note         = {OpenAI API documentation, accessed 2026-04-10}
}

@misc{yu2026chipbenchnextstepbenchmarkevaluating,
      title={ChipBench: A Next-Step Benchmark for Evaluating LLM Performance in AI-Aided Chip Design}, 
      author={Zhongkai Yu and Chenyang Zhou and Yichen Lin and Hejia Zhang and Haotian Ye and Junxia Cui and Zaifeng Pan and Jishen Zhao and Yufei Ding},
      year={2026},
      eprint={2601.21448},
      archivePrefix={arXiv},
      primaryClass={cs.AI},
      url={https://arxiv.org/abs/2601.21448}, 
}

@article{pan2025llm_eda_survey,
  author    = {Jingyu Pan and Guanglei Zhou and Chen-Chia Chang and Isaac Jacobson and Jiang Hu and Yiran Chen},
  title     = {A Survey of Research in Large Language Models for Electronic Design Automation},
  journal   = {ACM Transactions on Design Automation of Electronic Systems},
  volume    = {30},
  number    = {3},
  pages     = {34:1--34:21},
  year      = {2025},
  doi       = {10.1145/3715324}
}

@article{nane2016hls_survey,
  author    = {Razvan Nane and Vlad-Mihai Sima and Christian Pilato and Jongsok Choi and Blair Fort and Andrew Canis and Yu Ting Chen and Hsuan Hsiao and Stephen Brown and Fabrizio Ferrandi and Jason Anderson and Koen Bertels},
  title     = {A Survey and Evaluation of FPGA High-Level Synthesis Tools},
  journal   = {IEEE Transactions on Computer-Aided Design of Integrated Circuits and Systems},
  volume    = {35},
  number    = {10},
  pages     = {1591--1604},
  year      = {2016},
  doi       = {10.1109/TCAD.2015.2513673}
}

@article{abdollahi2025hdv_llm_review,
  author    = {Meisam Abdollahi and Seyedeh Faegheh Yeganli and Mohammad (Amir) Baharloo and Amirali Baniasadi},
  title     = {Hardware Design and Verification with Large Language Models: A Scoping Review, Challenges, and Open Issues},
  journal   = {Electronics},
  volume    = {14},
  number    = {1},
  pages     = {120},
  year      = {2025},
  doi       = {10.3390/electronics14010120},
  url       = {https://www.mdpi.com/2079-9292/14/1/120}
}

@article{cong2011hls_fpga,
  author    = {Jason Cong and Bin Liu and Stephen Neuendorffer and Juanjo Noguera and Kees Vissers and Zhiru Zhang},
  title     = {High-Level Synthesis for {FPGAs}: From Prototyping to Deployment},
  journal   = {IEEE Transactions on Computer-Aided Design of Integrated Circuits and Systems},
  volume    = {30},
  number    = {4},
  pages     = {473--491},
  year      = {2011},
  doi       = {10.1109/TCAD.2011.2110592}
}

@inproceedings{rtllm,
  author    = {Yao Lu and Shang Liu and Qijun Zhang and Zhiyao Xie},
  title     = {RTLLM: An Open-Source Benchmark for Design RTL Generation with Large Language Model},
  booktitle = {2024 29th Asia and South Pacific Design Automation Conference (ASP-DAC)},
  pages     = {722--727},
  year      = {2024},
  doi       = {10.1109/ASP-DAC58780.2024.10473904}
}

@inproceedings{donlin2004tlm_flows,
  author    = {Adam Donlin},
  title     = {Transaction Level Modeling: Flows and Use Models},
  booktitle = {Proceedings of the International Conference on Hardware/Software Codesign and System Synthesis},
  pages     = {75--80},
  year      = {2004},
  doi       = {10.1109/CODESS.2004.240821}
}

@inproceedings{cai2003tlm_overview,
  author    = {Lukai Cai and Daniel Gajski},
  title     = {Transaction Level Modeling: An Overview},
  booktitle = {Proceedings of the First IEEE/ACM/IFIP International Conference on Hardware/Software Codesign and System Synthesis},
  pages     = {19--24},
  year      = {2003},
  doi       = {10.1109/CODESS.2003.1275250}
}

@ARTICLE{cordic,
  author={Verma, Anu and Kiyawat, Khyati and Das, Bishnu Prasad and Meher, Pramod Kumar},
  journal={IEEE Transactions on Very Large Scale Integration (VLSI) Systems}, 
  title={An Efficient Scaling-Free Folded Hyperbolic CORDIC Design Using a Novel Low-Complexity Power-of-2 Taylor Series Approximation}, 
  year={2023},
  volume={31},
  number={8},
  pages={1167-1177},
  doi={10.1109/TVLSI.2023.3281078}}

@article{shi2026cktevo,
  title={CktEvo: Repository-Level RTL Code Benchmark for Design Evolution},
  author={Shi, Zhengyuan and Wang, Jingxin and Cheng, Tairan and Xu, Changran and Qian, Weikang and Xu, Qiang},
  journal={arXiv preprint arXiv:2603.08718},
  year={2026}
}

@misc{synopsys_design_compiler,
  author       = {{Synopsys, Inc.}},
  title        = {{Design Compiler: Timing, Area, Power, \& Test Optimization}},
  year         = {2026},
  howpublished = {\url{https://www.synopsys.com/implementation-and-signoff/rtl-synthesis-test/design-compiler.html}},
  note         = {Accessed: 2026-04-14}
}

@inproceedings{tine2021vortex,
  author    = {Tine, Blaise and Yalamarthy, Krishna Praveen and Elsabbagh, Fares and Kim, Hyesoon},
  title     = {Vortex: Extending the {RISC-V} {ISA} for {GPGPU} and 3D-Graphics},
  booktitle = {Proceedings of the 54th Annual IEEE/ACM International Symposium on Microarchitecture},
  series    = {MICRO '21},
  year      = {2021},
  pages     = {754--766},
  publisher = {ACM},
  doi       = {10.1145/3466752.3480128}
}

@misc{yu2026horizon,
  author       = {Yu, Cunxi and Deng, Chenhui and Pinckney, Nathaniel and Khailany, Brucek},
  title        = {Agentic Hardware Design as Repository-Level Code Evolution},
  year         = {2026},
  eprint       = {2606.28279},
  archivePrefix = {arXiv},
  primaryClass = {cs.AR},
  doi          = {10.48550/arXiv.2606.28279}
}

@article{fang2025rtlcoder,
  author  = {Wenji Fang and Yao Lu and Shang Liu and Qijun Zhang and Ceyu Xu and Lisa Wu Wills and Hongce Zhang and Zhiyao Xie},
  title   = {{RTLCoder}: Fully Open-Source and Efficient {LLM}-Assisted {RTL} Code Generation Technique},
  journal = {IEEE Transactions on Computer-Aided Design of Integrated Circuits and Systems},
  volume  = {44},
  number  = {4},
  pages   = {1448--1461},
  year    = {2025}
}

@inproceedings{lattner2021mlir,
  author    = {Chris Lattner and Mehdi Amini and Uday Bondhugula and Albert Cohen and Andy Davis and Jacques Pienaar and River Riddle and Tatiana Shpeisman and Nicolas Vasilache and Oleksandr Zinenko},
  title     = {{MLIR}: Scaling Compiler Infrastructure for Domain Specific Computation},
  booktitle = {2021 IEEE/ACM International Symposium on Code Generation and Optimization (CGO)},
  pages     = {2--16},
  year      = {2021}
}

@inproceedings{eldridge2021mlir_hardware,
  author    = {Schuyler Eldridge and Prithayan Barua and Aliaksei Chapyzhenka and Adam Izraelevitz and Jack Koenig and Chris Lattner and Andrew Lenharth and George Leontiev and Fabian Schuiki and Ram Sunder and Andrew Young and Richard Xia},
  title     = {{MLIR} as Hardware Compiler Infrastructure},
  booktitle = {Workshop on Open-Source EDA Technology (WOSET)},
  year      = {2021},
  url       = {https://woset-workshop.github.io/PDFs/2021/a06.pdf}
}

@misc{yin2026cppl,
  author        = {Shuo Yin and Yihe Wang and Lancheng Zou and Xufeng Yao and Tinghuan Chen and Chen Bai and Zhengrong Wang and Tsung-Yi Ho and Bei Yu},
  title         = {{CPPL}: A Circuit Prompt Programming Language},
  year          = {2026},
  eprint        = {2605.17892},
  archivePrefix = {arXiv},
  primaryClass  = {cs.AR},
  url           = {https://arxiv.org/abs/2605.17892}
}

@misc{synopsys_formality,
  author       = {{Synopsys, Inc.}},
  title        = {{Formality Equivalence Checking}},
  year         = {2026},
  howpublished = {\url{https://www.synopsys.com/implementation-and-signoff/signoff/formality-equivalence-checking.html}},
  note         = {Accessed: 2026-07-16}
}

@inproceedings{tan2026autoverifix,
  author    = {Yan Tan and Xiangchen Meng and Zijun Jiang and Yangdi Lyu},
  title     = {{AutoVeriFix}: Automatically Correcting Errors and Enhancing Functional Correctness in {LLM}-Generated Verilog Code},
  booktitle = {2026 31st Asia and South Pacific Design Automation Conference (ASP-DAC)},
  pages     = {526--532},
  year      = {2026},
  doi       = {10.1109/ASP-DAC66049.2026.11420300}
}
\end{document}